\documentclass[onecolumn,12pt,nofootinbib]{revtex4-1}

\usepackage[english]{babel}
\usepackage{color}
\usepackage{amsmath}
\usepackage{physics}
\usepackage[hidelinks]{hyperref}
\usepackage{diagbox}
\usepackage{multirow}
\usepackage{amsfonts}
\usepackage{amssymb}
\usepackage{latexsym}
\usepackage{graphicx}
\usepackage{adjustbox}
\usepackage{xr}
\usepackage{siunitx}
\usepackage{threeparttable}
\usepackage{braket}
\usepackage[justification=centering]{caption}
\usepackage{float}
\usepackage{caption}
\usepackage{subcaption}
\usepackage{empheq}
\usepackage{array}
\usepackage{seqsplit}

\newcommand{\al}{\alpha}

\newcommand{\la}{\lambda}

\newcommand{\rar}{\rightarrow}

\begin{document}

\title{Comment on: \textit{ Uncommonly accurate energies for the general quartic oscillator}, Int. J. Quantum Chem., e26554 (2020), by P.~Okun and K.~Burke}
\date{\today}

\author{Alexander~V.~Turbiner}
\email{turbiner@nucleares.unam.mx}
\author{Juan Carlos~del~Valle}
\email{delvalle@correo.nucleares.unam.mx}
\affiliation{Instituto de Ciencias Nucleares, Universidad Nacional
Aut\'onoma de M\'exico, Apartado Postal 70-543, 04510 M\'exico,
D.F., Mexico{}}

\begin{abstract}
It is shown that for the one-dimensional quantum quartic anharmonic oscillator
the numerical results obtained by Okun-Burke in Ref.\,2 are easily reproduced and can be significantly improved in Lagrange Mesh Method (based on non-uniform lattice).
\end{abstract}

\maketitle

There is a common opinion among people working on one-dimensional quantum dynamics that
the spectra of eigenvalues of one-dimensional Schr\"odinger equation can be found numerically with {\it any} desirable accuracy (unlike eigenfunctions), see e.g. \cite{FMF} and references therein~\footnote{In particular, it is shown that in the so-called Riccati-Pade method one can easily reach several hundred figures in energies
for polynomial potentials. Hence, the use in the title of \cite{Burke} the wording {\it Uncommonly accurate energies \ldots} is misleading.}. The question is how much time is ready to be invested and what is the most economic way to do so \footnote{It is necessary to mention that high-accuracy calculations (say more than 10 figures) suffer from the presence of different corrections (relativistic, non-pointness of masses, thus, mass distribution effects etc) which has to be taken into account to get physically relevant results.}. Variational method is the most common candidate. In Ref.\cite{Burke} the simplest trial function in the form of linear superposition of the harmonic oscillator eigenfunctions, we denote it as $\Psi_{HO}$, was used for quartic anharmonic oscillator
\begin{equation*}
   -\frac{1}{2}\,\frac{d^2\Psi}{dx^2}\ +\ V(x)\,\Psi\ =\ E\,\Psi\ ,\qquad
   V(x)\ =\ -\frac{\la}{2}\,x^2\ +\ \frac{1}{4}\,x^4\ ,
\end{equation*}
to find energies at nine values of parameter $\la$, ranging from -1 to 16, for 20 lowest eigenvalues. It was obtained 41 figures in each energy \footnote{CPU time needed for calculations was not indicated.}. For this type of trial functions the variational energy (saying differently, the expectation value of the Hamiltonian) can be calculated analytically, which simplifies the process of minimization if it is needed.
However, it is well known that this type of trial functions has an intrinsic deficiency: it does {\it not} reproduce the correct asymptotic behavior of the exact eigenfunctions at large distances, which is $$\sim e^{- \frac{1}{3\sqrt{2}} |x|x^2}\ .$$ It leads to increasingly slow convergence with the increase of the number of correct figures in eigenvalues~
\footnote{Following numerous critical remarks by W.~Kutzelnigg we think that due to this reason (as well as impossibility to reproduce the cusp parameters correctly) the Gaussian orbitals were not used to get benchmark results for the helium atom, see \cite{Sch-NN-Kor}, contrary to the Coulomb orbitals.}. Furthermore, it gives wrong expectations values: for any $\al > 0$ the expectation value
\[
      < \exp{a x^{2+\al}} >|_{\Psi_{OB}}\ =\  \infty \ ,
\]
for $a > 0$, although for $\al < 1$ this expectation should be finite. This deficiency stems from the fact that the trial potential
\[
V_{trial}\ =\ \frac{1}{\Psi_{HO}}\,\bigg(\frac{d^2}{dx^2} \Psi_{HO}\bigg)\ ,
\]
\pagebreak
behaves at large distances like $\sim x^2$, while the perturbation potential $(V-V_{trial})$ grows like $\sim x^4$, hence, dominant with respect to $V_{trial}$ and the perturbation theory in deviation $(V-V_{trial})$ is divergent. It implies that the first correction to variational energy does {\it not} provide the correct estimate of the accuracy in variational energy (for discussion see \cite{Turbiner:1980-4}).

The goal of this Comment is to demonstrate that in Lagrange Mesh Method \cite{Baye:2015} (and references therein), based on non-uniform discretization following the zeroes of Hermite polynomials, the results obtained in \cite{Burke} can be easily reproduced - it is a matter of seconds in CPU time - and easily overtaken reaching the accuracy $\gtrsim$ 200 figures in a matter of minutes in CPU time in standard laptop with Mathematica code. Present authors have definite doubts that similar accuracies can be reached in the method used in \cite{Burke} in comparable times.

To simplify presentation let us introduce the following notation
\begin{equation*}
 \underset{Y_Z}{X}\ :\qquad X\ =\  \text{Digit}\ ,\qquad Y\ =\  \text{Mesh Points}\ ,
 \qquad Z\ =\  \text{Decimal Place of $X$} \quad ,
\end{equation*}
all marked by red color. Digit $X$ indicates the maximal digit in energy which is reproduced with number of mesh points $Y$ \footnote{Accuracy of 300 digits in definition of mesh points used.  In Mathematica, \textbf{WorkingPrecision}$\rar$300. }. Maximal accuracy of the energy is calculated with 2000 mesh points and checked with 2020 mesh points. Maximal digit, reached in calculations by Okun \& Burke (O\&B), see \cite{Burke}, is marked by bold, see below.

Concrete calculations we made for $\la=-1$ (single well potential case) for the ground state energy and for {$\la=1,16$} (double well potential case) for the ground state $(\la=1,16)$ and for 19th excited state {$(\la=1)$} energies. Computations were carried out in Mathematica-12 using iMac with 2.7 GHz Intel Core i5 with 8GB RAM.


\subsection*{Quartic Anharmonic Oscillator: $\lambda=-1$}

As for the ground state energy:

\begin{align*}
E_0\ =\  0.&620\,927\,0
\underset{\mathclap{\textcolor{red}{25{{}_{8}}}}}{\textcolor{red}{2}}
9
\,825\,7
\underset{\mathclap{\textcolor{red}{50{{}_{14}}}}}{\textcolor{red}{4}}
8\,660\,8
\underset{\mathclap{\textcolor{red}{75{{}_{20}}}}}{\textcolor{red}{5}}
8\,035\,
\underset{\mathclap{\textcolor{red}{100{{}_{25}}}}}{\textcolor{red}{7}}
32\,987\,12
\underset{\mathclap{\textcolor{red}{150{{}_{33}}}}}{\textcolor{red}{0}}
\,698\,200\,
\overset{\mathclap{\textbf{O\&B}}}{\textbf{\underline{0}}}
17\,2
\underset{\mathclap{\textcolor{red}{200{{}_{44}}}}}{\textcolor{red}{5}}
3\,619\,1
\underset{\mathclap{\textcolor{red}{250{{}_{50}}}}}{\textcolor{red}{3}}
8\,982\,542\,3
\underset{\mathclap{\textcolor{red}{300{{}_{59}}}}}{\textcolor{red}{6}}
7\,325\\
&062\,962\,74
\underset{\mathclap{\textcolor{red}{400{{}_{72}}}}}{\textcolor{red}{8}}
\,188\,768\,883\,979\,39
\underset{\mathclap{\textcolor{red}{500{{}_{87}}}}}{\textcolor{red}{1}}
\,351\,303\,479\,456\,083\,601\,618\,760\,073\,476\,624\,891\,085\\
&768\,308\,099\,065\,938\,402\,
\underset{\mathclap{\textcolor{red}{1000{{}_{145}}}}}{\textcolor{red}{5}}
80\,084\,530\,397\,024\,737\,474\,347\,663\,406\,954\,493\,075\,566\,093\\
&052\,396\,859\,302\,472\,486\,392\,601\,975\,136\,357\,293\,108\,871\,529\,43
\underset{\mathclap{\textcolor{red}{1900{{}_{237}}}}}{\textcolor{red}{9}}
\,117\,092\,27
\underset{\mathclap{\textcolor{red}{2000_{{}_{246}}}}}{\textcolor{red}{5}}
\end{align*}

It is worth noting that the use of the 25 mesh points allows to get 9 figures (8 decimal digits (d.d.)) in $\sim 0.2$\,sec\,, see Table I. The result by [2] of 40 d.d. requires about 180 mesh points: it takes $\sim 4$\,sec\,, see Table I. For 2000 mesh points 246 d.d. are reached, it takes $\sim 35$\,min\,, see Table I. It was checked that similar accuracies with similar CPU times are reached for the first 80 eigenstates for both energies and node positions with similar CPU times. Let us emphasize that the rate of convergence is about 10-11 correct digits with increment of the number of mesh points in 100.

\subsection*{Double Well Potential: $\lambda=1$}

As for the ground state energy:

\begin{align*}
E_0\ =\ 0.&147\,235\,
\underset{\mathclap{\textcolor{red}{25_{{}_{7}}}}}{\textcolor{red}{1}}
40\,090\,0
\underset{\mathclap{\textcolor{red}{50_{{}_{14}}}}}{\textcolor{red}{3}}
5\,649\,
\underset{\mathclap{\textcolor{red}{75_{{}_{19}}}}}{\textcolor{red}{9}}
69\,12
\underset{\mathclap{\textcolor{red}{100_{{}_{24}}}}}{\textcolor{red}{4}}
\,897\,756\,466\,017\,325\,
\underset{\mathclap{\textcolor{red}{200_{{}_{40}}}}}
{\overset{\mathclap{\textbf{O\&B}}}{\textbf{\underline{\textcolor{red}{7}}}}}
55\,318\,874\,539\,254\,
\underset{\mathclap{\textcolor{red}{300_{{}_{55}}}}}{\textcolor{red}{9}}
92\,800\,263\\
             &120\,98
\underset{\mathclap{\textcolor{red}{400_{{}_{69}}}}}{\textcolor{red}{1}}
          \,357\,377\,138\,079\,999\,
\underset{\mathclap{\textcolor{red}{500_{{}_{85}}}}}{\textcolor{red}{9}}
82\,297\,179\,296\,021\,890\,349\,762\,419\,246\,096\,725\,129\,055\\
             &929\,407\,582\,589\,8
\underset{\mathclap{\textcolor{red}{1000_{{}_{140}}}}}{\textcolor{red}{4}}
5\,981\,955\,896\,482\,547\,701\,719\,569\,216\,318\,159\,102\,998\,544\,658\,831\\
             &561\,784\,177\,853\,835\,264\,833\,386\,947\,372\,208\,934\,630\,0
             \underset{\mathclap{\textcolor{red}{1900_{{}_{230}}}}}{\textcolor{red}{1}}
             2\,928\,564\,7
  \underset{\mathclap{\textcolor{red}{2000_{{}_{239}}}}}{\textcolor{red}{8}}
\end{align*}

\noindent
It is worth noting that the use of the 25 mesh points allows to get 8 figures (7 d.d.) in $\sim 0.2$\,sec\,, see Table II. The result by [2] of 40 d.d. requires about 200 mesh points: it takes $\sim 16$\,sec\,, see Table II. For 2000 mesh points 239 d.d. are reached, it takes $\sim 41$\,min\,, see Table II.

While as for the 19th excited state energy:

\begin{align*}
E_{19}\ =\  42.&
\underset{\mathclap{\textcolor{red}{50_{{}_{1}}}}}{\textcolor{red}{3}}
87\,460\,3
\underset{\mathclap{\textcolor{red}{100_{{}_{8}}}}}{\textcolor{red}{9}}
8\,659\,97
\underset{\mathclap{\textcolor{red}{150{{}_{15}}}}}{\textcolor{red}{6}}
\,360\,748\,
\underset{\mathclap{\textcolor{red}{200_{{}_{22}}}}}{\textcolor{red}{4}}
60\,339\,151\,
\underset{\mathclap{\textcolor{red}{250_{{}_{31}}}}}{\textcolor{red}{3}}
40\,412\,
\underset{\mathclap{\textcolor{red}{300_{{}_{37}}}}}{\textcolor{red}{5}}
21\
\overset{\mathclap{\textbf{O\&B}}}{\textbf{\underline{4}}}
74\,939\,156\,83
\underset{\mathclap{\textcolor{red}{400{{}_{51}}}}}{\textcolor{red}{5}}
\,342\,873\,143\,475\\
&\underset{\mathclap{\textcolor{red}{500_{{}_{64}}}}}{\textcolor{red}{8}}
34\,442\,346\,776\,630\,858\,786\,014\,482\,728\,909\,852\,009\,515\,813\,795\,919\,753\,312\,
\underset{\mathclap{\textcolor{red}{1000_{{}_{121}}}}}{\textcolor{red}{2}}
44\,257\\
&330\,182\,061\,161\,689\,338\,128\,957\,261\,369\,362\,484\,027\,548\,806\,789\,503\,865\,374\,787\,715\\
&466\,718\,578\,447\,536\,669\,
\underset{\mathclap{\textcolor{red}{1900{{}_{208}}}}}{\textcolor{red}{8}}
10\,863\,978\,04
\underset{\mathclap{\textcolor{red}{2000{{}_{219}}}}}{\textcolor{red}{1}}
\end{align*}

It is worth noting that in this case the use of 50 mesh points allows to get 3 figures (1 d.d.) in $\sim 4.$\,sec\,, see Table II.  The result by [2] of 40 d.d. requires about 330 mesh points: it takes $\sim 31$\,sec\,, see Table II. For 2000 mesh points 219 d.d. are reached, it takes $\sim 41$\,min\,, see Table II.

\subsection*{{Double Well Potential: $\lambda=16$}}
{It can be called the extreme double well potential case. As for the ground state energy:
	\begin{align*}
	E_0\ =\  -61.&1
	\underset{\mathclap{\textcolor{red}{25_{{}_{2}}}}}{\textcolor{red}{8}}
	7\,397\,
	\underset{\mathclap{\textcolor{red}{50_{{}_{7}}}}}{\textcolor{red}{6}}
	09\,72
	\underset{\mathclap{\textcolor{red}{100_{{}_{12}}}}}{\textcolor{red}{3}}
	\,934\,70
	\underset{\mathclap{\textcolor{red}{150_{{}_{18}}}}}{\textcolor{red}{4}}
	\,051\,95
	\underset{\mathclap{\textcolor{red}{200_{{}_{24}}}}}{\textcolor{red}{1}}
	\,487\,8
	\underset{\mathclap{\textcolor{red}{250_{{}_{29}}}}}{\textcolor{red}{3}}
	7\,640\,84
	\underset{\mathclap{\textcolor{red}{300_{{}_{36}}}}}{\textcolor{red}{7}}
	\,511\,
	\overset{\mathclap{\textbf{O\&B}}}{\textbf{\underline{0}}}
	44\,866\,2
	\underset{\mathclap{\textcolor{red}{400_{{}_{47}}}}}{\textcolor{red}{6}}
	5\,399\,919\,578\,
	\underset{\mathclap{\textcolor{red}{500_{{}_{58}}}}}{\textcolor{red}{8}}
	34\,499\\
	&630\,898\,026\,753\,728\,525\,948\,951\,003\,309\,559\,623\,352\,261\,458\,
	\underset{\mathclap{\textcolor{red}{1000_{{}_{109}}}}}{\textcolor{red}{5}}
	67\,996\,340\,964\,347\,302\\
	&074\,068\,801\,017\,081\,360\,119\,109\,362\,199\,469\,453\,000\,146\,444\,413\,730\,116\,152\,941\,798\\
	&
	\underset{\mathclap{\textcolor{red}{1900_{{}_{190}}}}}{\textcolor{red}{9}}
	36\,942\,63
	\underset{\mathclap{\textcolor{red}{2000_{{}_{198}}}}}{\textcolor{red}{7}}
	\end{align*}}
{
	In this case, by using 25 mesh points it allows to get 4 figures (2 d.d.) in $\sim 0.3$\,sec\,, see Table III.  The result by [2] of 40 d.d. requires about 330 mesh points: it takes $\sim 15$\,sec\,, see Table III. For 2000 mesh points 198 d.d. are reached, it takes $\sim 34$\,min\,, see Table III.}

\subsection*{Running Time}

\begin{table}[h]
	\caption{CPU time (R. Time) needed to compute the ground state energy at $\la=-1$ in
     Lagrange-Mesh Method {\it versus} the number of mesh points (Mesh P.). }
	{\setlength{\tabcolsep}{0.4cm}	
		\begin{tabular}{cccc}
			\hline
			\hline
			Mesh P. & R. Time & Mesh P.& R. Time \\
			\hline
			25   &  0.23 s        & 250     &  7.20 s       \\
			50   &  0.38 s        & 300     &  11.25 s       \\
			75   &  0.64 s        & 400     &  23.05 s       \\
			100  & 1.06 s        & 500     &  41.77 s        \\
			150  & 2.13 s        & 1000   & 4.85 min      \\
			200  & 4.15 s        &  2000  & 35.85 min \\
			\hline
			\hline		
		\end{tabular}
	}
\end{table}

\begin{table}[h]
	\caption{CPU time (R. Time) needed to compute the energy of any of the first 20 low-lying
    states at $\la=1$ in Lagrange-Mesh Method \textit{versus}  the number of mesh points (Mesh P.).} 
	{\setlength{\tabcolsep}{0.4cm}	
		\begin{tabular}{cccc}
			\hline
			\hline
			Mesh P. & R. Time & Mesh P.& R. Time \\
			\hline
			25   &   0.22 s        & 250    &   21.42 s       \\
			50   &   4.12 s        & 300    &  28.16 s       \\
			75   &    7.50 s       & 400    &   47.53 s       \\
			100  &  9.55 s        & 500    &  1.20 min        \\
			150  &   12.32 s     & 1000  & 6.10 min      \\
			200  &  16.09 s      & 2000  & 41.06 min \\
			\hline
			\hline		
		\end{tabular}
	}
\end{table}
\begin{table}[h]
	{
		\caption{CPU time (R. Time) needed to compute the ground state energy at $\la=16$ in Lagrange-Mesh Method \textit{versus} the number of mesh points (Mesh P.).}
		{\setlength{\tabcolsep}{0.4cm}	
			\begin{tabular}{cccc}
				\hline
				\hline
				Mesh P. & R. Time & Mesh P.& R. Time \\
				\hline
				25   &  0.27 s        & 250     &  8.59 s       \\
				50   &  0.48 s        & 300     &  12.76 s       \\
				75   &  1.65 s        & 400     &  24.73 s       \\
				100  & 1.06 s        & 500     &  42.86 s        \\
				150  & 3.14 s        & 1000   & 4.76 min      \\
				200  & 5.13 s        &  2000  & 34.06 min \\
				\hline
				\hline		
			\end{tabular}
		}
	}
\end{table}
\newpage
As the conclusion we have to mention that there are no real obstacles to increase the number of mesh points in the Lagrange Mesh Method further, hence, to increase accuracy in eigenvalues of the quartic oscillator.
{For $-1\leq\la\leq16$, the CPU time needed to calculate de ground state energy with given accuracy is basically  independent on $\la$.}
This method can be easily applied to any polynomial potential, which has the discrete spectra, for calculation of the eigenvalues. It was checked for the general radial anharmonic oscillator \cite{delValle} leading to the benchmark results for cubic, quartic and sextic radial oscillators. Furthermore, the Lagrange Mesh Method, which uses non-uniform lattice, allows to get easily the highly-accurate results comparable (or better) the existing benchmark results for the low-lying states of hydrogen atom in a constant uniform magnetic field of arbitrary strength \cite{delValle-Zeeman}.

\vspace{-5mm}
\section*{Acknowledgments}
\vspace{-3mm}

This work is partially supported by CONACyT grant A1-S-17364 and DGAPA grant IN113819 (Mexico).

\end{document}